\documentclass[10pt,twocolumn]{IEEEtran} 
\usepackage{cite}
\usepackage{amsmath,amssymb,amsfonts}
\usepackage{algorithmic}
\usepackage{graphicx}
\usepackage{textcomp}
\usepackage{xcolor}
\usepackage{todonotes}
\usepackage{enumerate}
\usepackage{tabularx}
\usepackage{multirow}
\usepackage{booktabs}
\usepackage{paralist}
\usepackage[braket, qm]{qcircuit}

\usepackage{xcolor}
\usepackage{tikzscale}
\usepackage{tikz}
\usetikzlibrary{shapes,arrows,decorations.pathreplacing,calc,arrows.meta,automata,shapes}
\usetikzlibrary{arrows,automata}
\usetikzlibrary{decorations.pathmorphing}
\usetikzlibrary{angles,quotes}
\usetikzlibrary{positioning}
\usetikzlibrary{fit}
\usetikzlibrary{mindmap, trees}
\usetikzlibrary{backgrounds}
\usetikzlibrary{intersections}
\usetikzlibrary{hobby,decorations.markings}
\usetikzlibrary{snakes}
\usepackage{pgfplots}
\usepgfplotslibrary{groupplots,units}
\usepackage{pgfplotstable}
\usepackage{tikz-qtree}
  \pgfdeclarelayer{background1}
  \pgfdeclarelayer{background2}
  \pgfsetlayers{background2,background1,main}
  
\usepackage{cleveref}
\crefname{section}{Section}{Sections}
\Crefname{section}{Section}{Sections}
\crefname{table}{\bf{Table}}{\bf{Table}}
\Crefname{table}{\bf{Table}}{\bf{Table}}
\crefname{figure}{\bf{Figure}}{\bf{Figure}}
\Crefname{figure}{\bf{Figure}}{\bf{Figures}} 
\crefname{definition}{Definition}{Definitions}
\crefname{equation}{}{}
\Crefname{equation}{Equation}{Equations}

\def\BibTeX{{\rm B\kern-.05em{\sc i\kern-.025em b}\kern-.08em
    T\kern-.1667em\lower.7ex\hbox{E}\kern-.125emX}}
\begin{document}

\title{Integration and Evaluation of Quantum Accelerators\\ for Data-Driven User Functions}

\author{\large Thomas Hubregtsen~$^{1,2}$,
Christoph Segler~$^{1,3}$,
Josef Pichlmeier~$^{2}$, \\
Aritra Sarkar~$^{2}$,
Thomas Gabor~$^{4}$, and 
Koen Bertels~$^{2}$
\\$^{1}$ BMW Group Research, New Technologies, Innovations, Garching bei
M\"unchen, Germany
\\$^{2}$ Delft University of Technology, Delft, Netherlands 
\\$^{3}$ Technical University of Munich, Department of Informatics,
Garching bei M\"unchen, Germany
\\$^{4}$ LMU Munich, M\"unchen, Germany 
}

\maketitle


\begin{abstract}

Quantum computers hold great promise for accelerating computationally challenging algorithms on noisy intermediate-scale quantum (NISQ) devices in the upcoming years. Much attention of the current research is directed towards algorithmic research on artificial data that is disconnected from live systems, such as optimization of systems or training of learning algorithms. In this paper we investigate the integration of quantum systems into industry-grade system architectures. In this work we propose a system architecture for the integration of quantum accelerators. In order to evaluate our proposed system architecture we investigated various data-driven functions for various accelerators, including a classical system, a gate-based quantum accelerator and a quantum annealer. The data-driven function predict user preference and is trained on real-world data. This work also includes an evaluation of the quantum enhanced kernel, that previously was only evaluated on artificial data. In our evaluation, we showed that the quantum-enhanced kernel performs at least equally well to a classical state-of-the-art kernel when simulated. We also showed a low reduction in accuracy and latency numbers within acceptable bounds when running on the gate-based IBM quantum accelerator. 
We therefore conclude it is feasible to integrate NISQ-era devices in industry-grade system architectures in preparation for future advancements in quantum hardware.

\end{abstract}


\section{\bf{Introduction}}
The industry today is facing major problems that are being solved by digital solutions. Optimization problems increase efficiency, data-driven functionalities provide forecasts and personalization; simulations provide a way to explore new compounds at scales previously unseen. 
Still, a major holdback is in the amount of data that can be processed. Just as with the advancement of deep neural networks through GPU acceleration, we are about to enter a new era of computation: the era of quantum computation. 
Certain algorithms that were previously thought to be inefficient in the generalized case, due to their computational complexity being labeled as non-deterministic polynomial-time (NP), are now becoming amenable to the quantum-efficient class called bounded quantum polynomial-time (BQP).
Furthermore, extensive research is being performed into the hypothesis that quantum systems can represent probability distributions that can not be matched efficiently on classical systems~\cite{aaronson2017complexity}, thereby promising greater expressive power. Algorithms already exist, such as the prime-factorization algorithm~\cite{th_shor} that threatens current-day security, and real-world experimental quantum systems are now accessible through the cloud. 

In this paper, a system architecture for the integration of quantum systems in industry-grade environments is proposed. The components and considerations of such an architecture are investigated. A working proof of concept is built to evaluate the performance in terms of timing and accuracy. For this, we implement and train a data-driven functionWe also evaluate the quantum-enhanced kernel proposed by Havlicek et~al.~\cite{Havlicek_2018} on real-world data, which previously was only tested on artificial data.

For this work, the following research questions are posed:
\begin{enumerate}[\bf RQ1]
    \item How to integrate quantum systems into an industry-grade system architecture?
    \item How do current-day state-of-the-art quantum accelerators perform in the production environment of real-world industry-grade data-centers?
    \item How does the quantum-enhanced kernel perform on real-world data?
\end{enumerate}

This paper provides the following scientific contributions, as well as the following contributions to the current state of practice:
\begin{enumerate}[\bf (i)]
    \item Proposed system architecture for integration of various quantum systems.
    \item Evaluation of quantum accelerators from a software and hardware perspective in a real-world scenario.
    \item Evaluation of the quantum-enhanced kernel on real-world data.
\end{enumerate}

\section{\bf{Related work}} \label{sec:related-work}
\subsection{\bf{Industry use-cases}}
The introduction of the D-Wave System's Quantum Annealer has inspired many scientists to implement real-world use-cases on an experimental quantum device~\cite{Yongcheng_2019,Kazuki_2019,Ors_2019}.
For instance, the applicability of the D-Wave 2X for traffic flow optimization~\cite{Neukart_Traffic} based on real-world data has been investigated.
In their work, the underlying optimization problem was defined, such that the lowest energy solution minimizes the number of cars that take the same route. Nevertheless, the experiment was performed on a small subset of the original problem.
Similarly, the D-Wave 2000Q has been used to approximate solutions for the flight gate assignment problem~\cite{DLR_Min_Transit}.
An optimal solution of the derived Quadratic Unconstrained Binary Optimization (QUBO) problem would minimize the transit time for passengers at an airport.
However, due to the limited problem size that can be processed with the D-Wave Quantum Annealer, only special subsets derived during pre-processing of the original problem have been used for the test. 

In contrast to the listed work, which takes a large problem and cuts out a subspace, we have selected a use-case that fits the current state-of-the-art quantum hardware.
We applied proven methods for reducing the number of calls, such as feature selection and cross-compilation. 
Furthermore, we proposed a system architecture and evaluated metrics relevant to the integration of this hardware, such as latency and queuing times.

\subsection{\bf{Quantum computing stack}}
To use a quantum processing unit~(QPU) for a particular application, there are many encapsulating layers that bridge the physical platform with the algorithmic logic~\cite{bertels2019quantum}.
These include expressing the application as an algorithm~\cite{sarkar2019algorithm} consisting of both classical and quantum kernels.
Typically the QPU is accessed as an accelerator from the host central processing unit~(CPU), with specific logic offloaded when quantum phenomena can be harnessed (e.g., superposition, entanglement, interference) to gain a computational advantage.
The quantum kernels in high-level logic are translated to assembly-level instructions for a gate-based quantum computer by the compiler~\cite{khammassi2018cqasm}.
This can be directly executed on a classical quantum-circuit simulator for functional verification~\cite{khammassi2017qx}.
In order to reach real QPU specifications (e.g., supported gateset, qubit connectivity, fidelity), we take into consideration further hardware constraints such as gate decomposition, scheduling, mapping, routing, and error-correction coding and apply them on the logic.
The resulting QPU-dependent assembly~\cite{fu2019eqasm} is then passed to the micro-architecture layer~\cite{riesebos2019quantum,fu2019control}.
This takes care of the precise instruction issue timing schedule for the analog waveform generators connected to the qubit control lines of the QPU chip.
Other groups~\cite{van2013blueprint, chong2018closing, martonosi2019next} have also pursued a system's view in the quantum hardware-software co-design for scaling up and encapsulating the QPU in a device agnostic accelerator framework.
Such a full-stack approach is crucial for large-scale simulators~\cite{huang2019alibaba} and both quantum accelerators based on annealers~\cite{fingerhuth2018open} and gate-based models~\cite{mckay2018qiskit}.

\section{\bf{Approach}} \label{sec:approach}
\begin{figure*}[h!]
	\centering
	\resizebox{1.4\columnwidth}{!}{%

\begin{tikzpicture}[
	node distance=3em,
	block/.style={rectangle, draw, fill=white, minimum width=3em, text width=3em, text centered, rounded corners=1pt,font=\small, inner sep = 0pt, minimum height=3em},
	blocksm/.style={rectangle, draw, fill=white, minimum width=4em, text width=4em, text centered, rounded corners=1pt,font=\small, inner sep = 0pt, minimum height=1.5em},
	blockbig/.style={rectangle, draw, fill=white, minimum width=4em, text width=4em, text centered, rounded corners=1pt,font=\small, inner sep = 0pt, minimum height=4em},
	ghost/.style={coordinate},
	api/.style={rectangle, draw, fill=white, minimum width=1.5em, text width=1.5em, text centered, rounded corners=1pt,font=\small, inner sep = 0pt, minimum height=4em},
	database/.style={cylinder, shape border rotate=90, aspect=0.3, draw, fill=white},
]

\node [block, minimum width=10em] (ecu1) {};
\node [blocksm, below right = 6em and 0em of ecu1.south west] (sensor1) {};
\node [blocksm, below left = 6em and 0em of ecu1.south east] (act1) {};
\draw [->] (sensor1) -- (ecu1.-153);
\draw [<-] (act1) -- (ecu1.-27);

\node [block, minimum width=10em, above left = 0.5em and 0.5em of ecu1.south east] (ecu2) {};
\node [blocksm, below left = 6em and 0em of ecu2.south east] (act2) {};
\node [blocksm, below right = 6em and 0em of ecu2.south west] (sensor2) {};
\draw [->] (sensor2) -- (ecu2.-153);
\draw [<-] (act2) -- (ecu2.-27);

\node [block, minimum width=10em, above left = 0.5em and 0.5em of ecu2.south east] (ecu3) {ECU};
\node [blocksm, below right = 6em and 0em of ecu3.south west] (sensor3) {Sensors};
\node [blocksm, below left = 6em and 0em of ecu3.south east] (act3) {Actuators};
\draw [->] (sensor3) -- (ecu3.-153);
\draw [<-] (act3) -- (ecu3.-27);

\node [blockbig, right = 2em of ecu2] (link) {Gate-\\way};
\draw [->] (ecu3) -- (link.165);
\draw [->] (ecu2) -- (link.177);
\draw [->] (ecu1) -- (link.192);

\node [blockbig, below right = -2.5em and 0.5em of link, dashed, minimum height=2.5em, minimum width=4.5em] (pre-veh) {Pre-\\processing};
\draw [->, dashed] (link.-20) -- (pre-veh);


\node [api, right = 7em of link] (api1) {\rotatebox{90}{API}};
\draw [->] (link.30) -- (api1.124);

\node [api, below right = 1em and 7em of link] (api2) {\rotatebox{90}{API}};

\node [blockbig, right = 0.5em of api1, minimum width=4.5em] (pre-back) {Pre-\\processing};
\node [database, right = 1em of pre-back] (db) {Data};
\node [blockbig, right = 1em of db] (fs) {Feature\\Selection};
\draw [->] (api1) -- (pre-back);
\draw [->] (pre-back) -- (db);
\draw [->] (db) -- (fs);

\node [api, below right = -2.1em and 1.5em of fs] (apitrain) {\rotatebox{90}{API}};
\node [blockbig, right = 0.75em of apitrain, minimum width=4.8em] (mltrain) {Data\\Driven\\Function};
\node [above = 0.25em of mltrain, shift={(-1em,0em)}] (texttrain) {Accelerator};
\draw [->] (db) -- (db |- apitrain.230) -- (apitrain.230);
\draw [<->] (apitrain) -- (mltrain);
\draw [->, densely dashed, >=stealth] (fs.-120) -- (fs.-120 |- apitrain.230);

\node [api, below = 6em of apitrain] (apiinfer) {\rotatebox{90}{API}};
\node [blockbig, below = 6em of mltrain, minimum width=4.8em] (mlinfer) {Data\\Driven\\Function};
\node [above = 0.25em of mlinfer, shift={(-1em,0em)}] (textinfer) {Accelerator};
\draw [->] (pre-back) -- (pre-back |- apiinfer.130) -- (apiinfer.130);
\draw [<-] (api2) -- (api2 |- apiinfer.220) -- (apiinfer.220);
\draw [<->] (apiinfer) -- (mlinfer);
\draw [->, densely dashed, >=stealth] (fs.-60) -- (fs.-60 |- apiinfer.130);

\node [blocksm, below = 7em of pre-veh.west, minimum width=4.8em] (apiveh) {API};
\node [blockbig,below = 0.75em of apiveh, minimum width=4.8em] (mlveh) {Data\\Driven\\Function};
\node [left = 0.25em of mlveh, shift={(0em,1em)}] (textinferveh) {\rotatebox{90}{Accelerator}};
\draw [<->] (apiveh) -- (mlveh);
\draw [<-, dashed] (apiveh.40) -- (apiveh.40 |- api2.-140) -- (pre-veh |- api2.-140) -- (pre-veh);
\draw [->, dashed] (apiveh.140) -- (apiveh.140 |- api2.-140) -- (ecu1.-20 |- api2.-140) -- (ecu1.-20);

\draw [->] (api2.140) -- (ecu1.-18 |- api2.140) -- (ecu1.-18);

\node [above left = 1.2em and -2.5em of ecu1.north west] (textveh) {Vehicle};
\node [shift={(2.5em,0em)}] at (api1 |- textveh) (textbackend) {OEM-Backend};
\node [above = 0.5em of texttrain] (cloudtrain) {Cloud Service};
\node [above = 0.5em of textinfer] (cloudinfer) {Cloud Service};

\begin{pgfonlayer}{background1}
	\node[fill=green!10, rounded corners=1pt, draw, dashed, fit=(apitrain)(mltrain)(texttrain), shift={(1em,-1em)}] (acc-train1) {};
	\node[fill=green!10, rounded corners=1pt, draw, dashed, fit=(apitrain)(mltrain)(texttrain), shift={(0.5em,-0.5em)}] (acc-train2) {};
	\node[fill=green!10, rounded corners=1pt, draw, dashed, fit=(apitrain)(mltrain)(texttrain)] (acc-train3) {};
	
	\node[fill=red!10, rounded corners=1pt, draw, dashed, fit=(apiinfer)(mlinfer)(textinfer), shift={(1em,-1em)}] (acc-infer1) {};
	\node[fill=red!10, rounded corners=1pt, draw, dashed, fit=(apiinfer)(mlinfer)(textinfer), shift={(0.5em,-0.5em)}] (acc-infer2) {};
	\node[fill=red!10, rounded corners=1pt, draw, dashed, fit=(apiinfer)(mlinfer)(textinfer)] (acc-infer3) {};
	\node[fill=red!10, rounded corners=1pt, draw, dashed, fit=(apiveh)(mlveh)(textinferveh)] (acc-veh) {};
\end{pgfonlayer}

\begin{pgfonlayer}{background2}
    \node[fill=black!10, rounded corners=1pt, draw, dashed, fit=(cloudinfer)(acc-infer1)(acc-infer2)(acc-infer3)] (lowestbg) {};
    \node[fill=black!10, rounded corners=1pt, draw, dashed, fit=(cloudtrain)(acc-train1)(acc-train2)(acc-train3)] (highestbg) {};
    
	\node[fill=black!10, rounded corners=1pt, draw, dashed, fit=(ecu3)(mlveh)(pre-veh)(textveh)(acc-veh)(ecu3 |- lowestbg.south)(ecu3 |- highestbg.north)] (bg-veh) {};
	
	\node[fill=black!10, rounded corners=1pt, draw, dashed, fit=(api1)(fs)(api2)(textbackend)(api2 |- lowestbg.south)(api2 |- highestbg.north)] (bg-back) {};

\end{pgfonlayer}

\end{tikzpicture}
}
	\caption{Proposed system architecture}
	\label{fig:architecture}  
	\vspace*{-1em}
\end{figure*}
An overview of the proposed system architecture is depicted in \cref{fig:architecture}. The architecture consists out of four major components:
\begin{inparaenum}[(i)]
\item The vehicle,
\item the backend of the original equipment manufacturer~(OEM) (i.e., the car maker),
\item the hardware accelerator for the training of the data-driven function (cf.~green boxes), and
\item the hardware accelerator for the deployment of the data-driven function (cf.~red boxes).
\end{inparaenum}
At first, the vehicle provides the data required to develop the data-driven function. The data from the vehicle is sent to the OEM's backend infrastructure, where the vehicle's data is received, preprocessed, and stored. For the training of the data-driven function, the training features are selected by feature selection. Only the selected subset of features is sent to the hardware accelerator for the training of the data-driven function hosted by a cloud service. The type of the accelerator can vary between different hardware designs, and in our case also include quantum accelerators. The developed data-driven function can then either be deployed on accelerating hardware in the backend or in the vehicle. 
\subsection{\bf{Vehicle}}
Most of the functionality in today's vehicles are realized as mechatronical systems. These are executed on electrical control units~(ECUs), of which current vehicles have up to 70~\cite{Broy.2006}. These ECUs are each connected to sensors and actuators in order to measure environmental as well as internal technical information as well as interact with their environment. Each of the ECUs provides their data to a gateway, which distributes the this data inside the vehicle or sends it to the backend of the OEM.
\subsection{\bf{OEM backend}}
The vehicle data arrives through an application programming interface (API) in the OEM's backend infrastructure. The data first goes through pre-processing, e.g. the data is cleaned (removal of invalid data, augmentation of context) and transformed (resampling, normalization). The data is then stored within the backend. 
To resolve the ``curse of dimensionality''~\cite{Bellman.1957}, posed by the high amount/dimensionality of vehicle signals, a feature selection component selects a subset of the most descriptive features/signals for the specific data-driven function. In our case, a supervised filter feature selection algorithm is applied in order to identify the subset of features. This feature selection algorithm calculates a score for each feature and rank them according to their score. The highest ranked features are used for the training of the data-driven function in the cloud service. This is ideal for quantum computation, as current state-of-the-art systems have a limited number of input features that can be processed.
\subsection{\bf{Cloud service}}
Data arrives in the accelerating cloud service at an API on a classical node. It is then forwarded to the accelerating hardware, where a data-driven function (often referred to as machine learning function) is trained on the pre-selected features and collected data. Each type of accelerating hardware has specific characteristics but often works together with a host CPU in a hybrid configuration. Typical hardware accelerators are GPUs, FPGAs, and ASICs. Roughly speaking, the GPU can perform many simple computations in parallel. The functions are given while the parameters can be changed. The FPGA can be reprogrammed to calculate complex functions, such as accelerating square roots efficiently. In this case, the developer defines both the function and the parameters. These functions can then be put together into an ASIC. The ASIC does not allow changes to the function, but can be further optimized both for performance and cost of mass production. In this case, only the parameters can be changed. 
   
Similarly, one can incorporate a quantum accelerator. One such accelerator heavily explored by industry is the special-purpose quantum annealer. This system allows for the acceleration of QUBO problems by providing a hardware platform and program where the user can modify the parameters. Other quantum accelerators are based on gate-based quantum systems. These are less mature technologies, but more generic and hold more promise. These systems allow the user to define both the program and the parameters along with the offered functionality of the hardware. Examples (in various stages of maturity) are the superconducting systems (e.g., Google, IBM~\cite{th_preskill_2019_nisq}), trapped-ion systems (e.g., IonQ~\cite{Wright_2019}), photonic systems (e.g., Xanadu~\cite{Pirandola_2018}) and topological systems (e.g., Microsoft~\cite{th_preskill_2019_nisq}). A mixture of quantum ideologies implemented on classical hardware also exist, in the form of quantum-inspired classical systems (Fujitsu~\cite{Aramon_2019}, Hitachi~\cite{Hitachi_2019}). It is currently an open question if these systems provide any advantage over conventional classical systems. 
   
After training, the model can be deployed. The deployment of the model for prediction would typically happen on a similar type of hardware accelerator, but one could consider functions that share a common mathematical formula to cross deploy on different hardware accelerators or simulators for training and prediction. For example, one could use a hybrid quantum-classical system to train the weights of a function, and deploy it on a classical system. In the case of the classical system, the prediction model could also be deployed directly in the vehicle architecture. When the data-driven function is deployed within the vehicle, the data pre-processing must also be performed onboard the vehicle (cf.~\cref{fig:architecture}).
\section{\bf{Evaluation}} \label{sec:evaluation}
\subsection{\bf{Test case}}
We will evaluate our architecture with a test in which we predict user preference, in particular the preference of the user regarding the use of seat-heating. 
The current state of the driver's seat-heating considering the states \emph{on} and \emph{off} is observed. The data-driven function is then trained on the historical data collected from the user in order to predict the seat-heating state. By this, a proactive seat-heating personalized on the user's behavior can be achieved.
\subsection{\bf{Data}}
Using one car, we collected 79 drives worth of raw data. The raw data was sampled at a of 5 seconds throughout the drive. In total, 116 features and 20458 samples were collected. We augmented 5 extra features, relating to the time of day and time elapsed since the start of the drive.
Every drive was additionally labeled with a majority vote to label the entire drive as seat heating on or off. Out of the 79 drives, 52 drives were performed with the seat heating off, 27 drives with the seat heating on. 
\subsection{\bf{Feature selection}}
In order to select the most relevant features as input for the data-driven function, the supervised filter feature selection algorithm \emph{Fisher Score}~\cite{Duda.2001} is run on the data. As a result the two most descriptive features are identified: the current outside temperature and the current temperature of the evaporator. Here, the evaporator is a part of the vehicle's Heating Ventilation Air Conditioning (HVAC). 
\subsection{\bf{Data-driven function}}
To predict the label for the chosen use-case, seat-heating, various functions are employed. The correlations in our data set allow for computationally less complex systems, such as decision trees, support vector machines (SVMs), and nearest neighbor. For our study, a SVM is deployed to find the best fit. A SVM takes the input data and maps it into a higher dimensional feature space, typically through a non-linear transformation called kernel function. Within this space, a linear separating hyperplane can be constructed to perform the classification of previously unseen data. SVMs are well-researched for classical hardware and are currently being explored for quantum accelerators. We investigate this on a classical CPU, a hybrid quantum/classical system involving a gate-based system from the IBM cloud~\cite{th_qiskit}, and a hybrid quantum/classical system involving a specific-purpose quantum annealer with 2000 qubits developed by D-Wave Systems~\cite{mcgeoch2019practical}. All classical calculation is performed on a single core of an Intel Xeon Gold 5222 CPU at 3.8 GHz.
\subsubsection{\bf{SVM on a classical CPU}}
We implement a SVM on a classical CPU. This is programmed in python and uses Scikit-learn\cite{th_pedregosa_2011_scikit-learn}. In particular, the fit and predict methods of \emph{sklearn.svm.SVC} with gamma set to auto are used. The following kernels are evaluated: linear, polynomial, radial basis function (RBF), and sigmoid. The model with the best performing kernel is used as a baseline model. This was the RBF kernel. 
\subsubsection{\bf{Gate-based quantum system}}
 
\begin{figure}[t]
    \includegraphics[width= 9 cm]{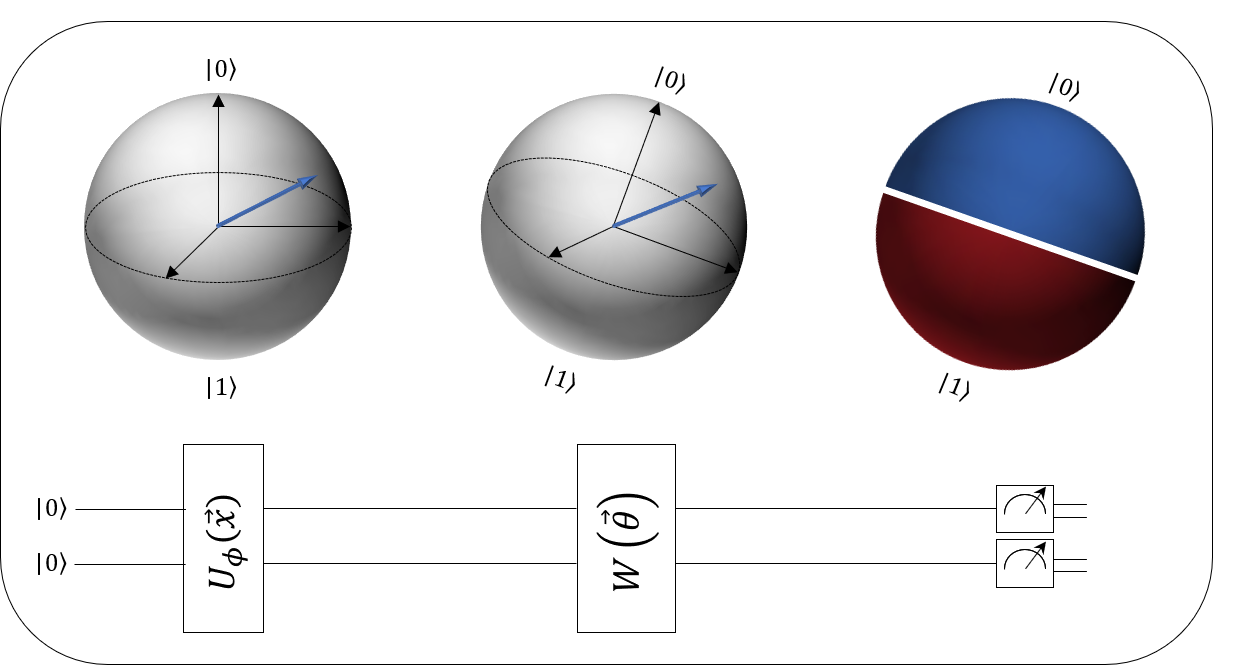}
    \vspace{-1.5em}
    \caption{Graphical Representation of the Quantum circuit}
	\label{fig:singleGate}  
	\vspace{-1em}
\end{figure}

\cref{fig:singleGate} illustrates the working principle of our implemented quantum support vector machine, such as proposed by Havlicek~et~al.~\cite{Havlicek_2018}. 
A simplified version of the quantum circuit is shown below. 
The abstract representation is shown at the top. Both are divided in three steps. 
First the embedding of the classical data as a quantum state through a quantum circuit $U_\Phi(\vec{x})$ is shown. 
This can be considered as the mapping of a feature vector $\vec{x}$ with dimension $N$ into a high dimensional space. 
This mapping should be non-linear, and hard to simulate with classical methods. 
Havlicek~et~al.~ used a circuit group for which the output distribution corresponds to that of an IQP circuit.
If the circuit is embedded multiple times, the resulting state is hard to estimate up to an additive polynomial small error. 
The system evolves according to the following operations:
\vspace{-0.2 cm}
\begin{equation}
\ket{\Phi(\vec{x})} = U_{\Phi(\vec{x})}~H^{\otimes n} U_{\Phi(\vec{x})}~H^{\otimes n} \ket{0}^{\otimes n}
\end{equation}
Since our input data is two dimensional, the interactions in the embedding circuit are at most quadratic. 
The operation $U_\Phi(\vec{x})$ can then be implemented according to following one- and two-body functions:
\vspace{-0.2 cm}
\begin{equation}
U_{\phi_{\{l,m\}}(\vec{x})} = e^{(i\phi_{\{l,m\}}(\vec{x})Z_kZ_l)}
\end{equation}
\vspace{-0.2 cm}
\vspace{-0.2 cm}
\begin{equation}
U_{\phi_{\{k\}}(\vec{x})} = e^{(i\phi_{\{k\}}(\vec{x})Z_k)}
\end{equation}
with $\phi_{\{1,2\}}(\vec{x}) = (\pi- x_1)(\pi - x_2)$ and $\phi_{\{k\}}(\vec{x}) = x_k$. 

In step 2, we use a variational circuit $W(\vec{\theta})$ in order to rotate the high-dimensional feature space, thereby shifting the decision boundary. 
This circuit consists of $X$ and $Y$ rotation unitaries and entangler gates to achieve a high degree of expressivity.
In the final step, the quantum state is measured and collapses into $N$ binary numbers. 
This bit string can be associated with an output label.
For one dimensional data, the measurement can be illustrated by slicing the Bloch sphere into two pieces which can be directly translated to label $0$ or $1$. For each data point $\vec{x}_n$, the process is repeated multiple times until a conclusive probability distribution is derived. \cite{Havlicek_2018}
 
The training of the SVM consists of finding the right parameters for the variational circuit. 
The quantum circuit, initialized with random values for $\theta$, is used to predict labels on the training data. 
Based on the resulting probability distribution and the known labels, a cost value is computed, and fed into the classical optimizer for proposing a new set of parameters. 
This is repeated until convergence is reached, or a maximum iteration limit is reached, whichever comes first. Predicting a label for a set of input values entails preparing the quantum state with the found parameters, and running the quantum circuit multiple times. The implementation of the circuit is available through the IBM qiskit library in a class called vqc~\cite{th_qiskit}. Both the training and the classification can independently be run. We train our system on a classical system using the IBM QASM simulator, and run our circuit in the IBM backend through a subscription service~\cite{Bartkiewicz_2019}.

\subsubsection{\bf{Specific-purpose quantum annealer}}
For the specific-purpose quantum annealer, we investigate the quantum accelerator built by D-Wave Systems, which specializes in approximating optimization problems which are given as an instance of a QUBO. In order to accelerate the construction of an SVM the problem needs to be translated into a QUBO, which was done according to the approach shown by Willsch et al.~\cite{willsch2019support}. The constructed QUBO can be written as a quadratic matrix of size $n \cdot k \times n \cdot k$ where $n$ is the amount of training data samples the SVM is based on and $k$ is the precision used for the discretization of the (originally real-valued) coefficients that define the SVM's decision boundary. 
\section{\bf{Results}} 
\begin{table*}[!t]
	\centering
	\caption{Evaluation results --- timing and accuracy of various methods and hardware accelerators}
	\label{tab:results}
	\scriptsize
	\begin{tabularx}{0.69\textwidth}{@{}ll|ll|lll@{}}
	\toprule 
	\hline
	\multicolumn{2}{c}{\textbf{Setup}} & 
	\multicolumn{2}{c}{\textbf{Training}} & 
	\multicolumn{3}{c}{\textbf{Validation}} \\\hline\hline
	$Method$ & $Kernel$ & $Hardware$ & $Time$ & $Hardware$ & $Time$ & $Accuracy$
	\\\hline
	\midrule
	Classical    & Rbf          & Classical & 74\hphantom{0} $\pm2$ ms & Classical & 485 $\pm5$\hphantom{0} ms           & 93.2$\pm0.0$\%     \\
	Gate-based   & Q. enhanced  & Simulator & 267 $\pm0$ m        & Simulator & 52\hphantom{0} $\pm0$\hphantom{0} s & 93.1$\pm0.1$\% \\
    Gate-based   & Q. enhanced  & Simulator & 267 $\pm0$ m        & Quantum   & 218 $\pm65$ m           & 91.2$\pm0.7$\% \\ \hline
    \bottomrule
    \end{tabularx}
\end{table*}
All evaluation runs are performed three times. 
The data was split into 90\% training data, 10\% test data, and 10\% validation data by splitting data based on the individual drives. It was made sure no samples of the same drive were present in both the training and validation data set. Hyper-parameters were trained on the training data and tested on the test data. Full experiments were performed by randomly sampling for the combined train and test data, and validating on the validation set. This resulted in 1813 samples, 1057 samples of which had the seat heating off, and 756 samples had the seat heating on. 
The models were verified on 100\% of the validation data, being 2327 samples. These 2327 samples contained 1830 samples with the seat heating off and 497 samples with the seat heating on. 
An overview of all results can be seen in \cref{tab:results}.
\subsection{\bf{Classical SVM}}
The training of the SVM on the classical system took between 72 to 76 milliseconds. The prediction on the test set took between 480 and 490 milliseconds. The RBF kernel was selected during tests on the test data. Using the RBF kernel, the accuracy achieved on the validation set was 93.2\%.
\subsection{\bf{Quantum Annealer}}
In the early phase of our work, the feasibility and potential of both the gate-based approach and the annealing-based approach were evaluated. However, the translation of the SVM construction to QUBO does not scale well and creates a graph that far exceeds the capabilities of the current generation quantum annealers. Tools exist to solve QUBOs that large classically or even with the support of the quantum annealer on specific sub-problems~\cite{booth2016abstractions}, but it was noticed that run-times jumped from sub-minute performance to over an hour. On a classical machine, the QUBO encoding has a quadratic disadvantage with respect to the number of data points used. Then solving QUBOs is NP-complete---while typical SVMs can be trained in cubic time---thus employing an algorithm (like TABU search) that is much more powerful and resource-intensive than necessary. These drawbacks cannot be overcome with the quantum annealer accelerating relatively small sub-problems compared to the full problem instance. For this reason, we choose to spend our remaining focus on the gate-based quantum accelerator. 
\subsection{\bf{Gate-based quantum accelerator}}
The training of the system on the classical hardware, consisting of tuning the parameters of the quantum circuit, took approximately 4.5 hours. 
The parameters found from training were used during validation.
Using the quantum simulator on classical hardware took 52 seconds and resulted in an accuracy between 93.1\% and 93.3\%. 
This experiment was also performed on the IBMQ quantum accelerator. This took between 2.5 hours and 5 hours, depending on the queue size. This resulted in an accuracy between 90.5\% and 91.9\%. The system reported error rates for its gates between $10^{-1}$ and $10^{-3}$. 
The evaluation of the 2327 samples on the IBMQ quantum accelerator involved running 2327 quantum circuits that needed to be repeated 1000 times each (called shots). Every call to the IBMQ API packed 75 circuits, and got queued for several seconds up to an hour. A total of 32 calls to the API were needed. Every pack of 75 circuits, ran 1000 times each, consumed 88 to 90 seconds of QPU time. Every single run of a circuit therefore completes in approximately 1.2 milliseconds. 
Sending data up for prediction from the vehicle to the cloud and back takes anywhere from sub-second time to several seconds, assuming the vehicle is in an area of connectivity. Sending a pro-active update is done once a minute. 
\section{\bf{Discussion}} \label{sec:discussion}
The accuracy of the quantum-enhanced kernel holds up in simulation to the accuracy of the classical RBF kernel with an accuracy of around 93\%. Even though the quantum system is subjective to error rates in the order of $10^{-1}$ to $10^{-3}$, whereas classical systems lie around $10^{-12}$, the performance decrease for the low-depth minimally entangled circuit is only 2\% points with a range of $\pm 1\%$ point. The authors of the method already showed with artificially created data that the algorithm works well with NISQ-era devices; with our findings, we show that this statement also holds on our real-world data. Future work would include investigating data with more features, multi-class labels, and various classes of use-cases, as well as exploring different feature maps. 

The quantum system, assuming an exclusive subscription to a machine, can classify a single sample with 1000 shots in 1.2 seconds. As the update loop in the car is on a minute-basis, we showed that a quantum system can be integrated for the prediction of in-vehicle systems using the architecture we proposed yielding near-similar accuracy as classical systems. 

Even though classical systems can perform classification on this small number of features quicker, with 32 microseconds per classification, and with similar accuracy, it is still expected that NISQ devices will eventually outperform classical systems for specific tasks similar to ours~\cite{th_preskill_2019_nisq}. First hints of nearing this computational barrier can already be seen, such as the recent experiment at Google~\cite{th_arute_2019_qs}.
\emph{Threats to the validity:}
The analysis was performed on data from one user on one use-case, being seat-heating. However, the focus of this paper is on the integration of the quantum system with regards to timing and reduction of accuracy due to decoherence. 

Additionally, the control parameters were not optimized. This included settings such as a number of shots and runs till convergence. This would be generated faster performance, but not change our argument, as the results are already within the minute-domain. 
\section{\bf{Conclusion}} \label{sec:conclusion}
Quantum computing is an emerging field with great potential. In this paper, we proposed an architecture integrating quantum accelerators in an industry-grade system. We implemented this architecture and evaluated the performance of various SVM implementations on real-world data for the prediction of the status of seat-heating. In our work, we were not searching for quantum advantage. The goal was to evaluate feasibility for the integration of NISQ devices in preparation for expected improvements in the next 5-10 years. These improvements would lay in more parallel processing power, and higher expressibility resulting in higher accuracy.   

The accuracy of the SVM with quantum-enhanced kernel on the gate-based quantum simulator performed similarly to a classical SVM with RBF kernel, both achieving an accuracy of 93\%. The same circuit run on the IBM quantum accelerator dropped 2\%-points in accuracy. This shows that the quantum-enhanced kernel, previously only tested on artificial data, also performs well on our real-world data. It also shows that our selected algorithm running on a NISQ-era device with high error rates can still perform close to its theoretical accuracy. 

The latency of the quantum system, assuming exclusive subscription to avoid lengthy and non-deterministic queues, can be parallelized to provide results in under a minute. This makes it feasible to predict for our seat-heating use-case, which requires updates every minute. 

Based on the low reduction in accuracy and latency numbers within acceptable bounds, we conclude it is feasible to integrate NISQ-era devices in an industry-grade system architecture in preparation for future hardware improvements.  

\bibliographystyle{unsrt} 
\bibliography{refs}

\begin{thebibliography}{10}

\bibitem{aaronson2017complexity}
Scott Aaronson and Lijie Chen.
\newblock Complexity-theoretic foundations of quantum supremacy experiments.
\newblock In {\em 32nd Computational Complexity Conference (CCC 2017)}. Schloss
  Dagstuhl-Leibniz-Zentrum fuer Informatik, 2017.

\bibitem{th_shor}
Peter~W. Shor.
\newblock Polynomial-time algorithms for prime factorization and discrete
  logarithms on a quantum computer.
\newblock {\em SIAM review 41.2 (1999): 303-332.}, 1999.

\bibitem{Havlicek_2018}
Vojtech~Havlicek et~al.
\newblock Supervised learning with quantum enhanced feature spaces.
\newblock 2018.

\bibitem{Yongcheng_2019}
Yongcheng~Ding et~al.
\newblock Towards prediction of financial crashes with a d-wave quantum
  computer, 2019.

\bibitem{Kazuki_2019}
Kazuki Ikeda, Yuma Nakamura, and Travis~S. Humble.
\newblock Application of quantum annealing to nurse scheduling problem, 2019.

\bibitem{Ors_2019}
Rom{\'{a}}n Or{\'{u}}s, Samuel Mugel, and Enrique Lizaso.
\newblock Quantum computing for finance: Overview and prospects.
\newblock {\em Reviews in Physics}, 4:100028, November 2019.

\bibitem{Neukart_Traffic}
Florian~Neukart et~al.
\newblock Traffic flow optimization using a quantum annealer, 2017.

\bibitem{DLR_Min_Transit}
Tobias Stollenwerk, Elisabeth Lobe, and Martin Jung.
\newblock Flight gate assignment with a quantum annealer, 2018.

\bibitem{bertels2019quantum}
K.~Bertels et~al.
\newblock Quantum computer architecture: Towards full-stack quantum
  accelerators.
\newblock {\em arXiv preprint arXiv:1903.09575}, 2019.

\bibitem{sarkar2019algorithm}
Aritra Sarkar, Zaid Al-Ars, Carmen~G Almudever, and Koen Bertels.
\newblock An algorithm for dna read alignment on quantum accelerators.
\newblock {\em arXiv preprint arXiv:1909.05563}, 2019.

\bibitem{khammassi2018cqasm}
Nader~Khammassi et~al.
\newblock cqasm v1. 0: Towards a common quantum assembly language.
\newblock {\em arXiv preprint arXiv:1805.09607}, 2018.

\bibitem{khammassi2017qx}
Nader~Khammassi et~al.
\newblock Qx: A high-performance quantum computer simulation platform.
\newblock In {\em 2017 Design, Automation \& Test in Europe Conference \&
  Exhibition (DATE)}, pages 464--469. IEEE, 2017.

\bibitem{fu2019eqasm}
Xiang~Fu et~al.
\newblock eqasm: An executable quantum instruction set architecture.
\newblock In {\em 2019 IEEE International Symposium on HPCA}, pages 224--237.
  IEEE, 2019.

\bibitem{riesebos2019quantum}
L~Riesebos et~al.
\newblock Quantum accelerated computer architectures.
\newblock In {\em IEEE ISCAS}, pages 1--4. IEEE, 2019.

\bibitem{fu2019control}
X~Fu, L~Lao, K~Bertels, and CG~Almudever.
\newblock A control microarchitecture for fault-tolerant quantum computing.
\newblock {\em Microprocessors and Microsystems}, 70:21--30, 2019.

\bibitem{van2013blueprint}
Rodney Van~Meter and Clare Horsman.
\newblock A blueprint for building a quantum computer.
\newblock {\em Communications of the ACM}, 56(10):84--93, 2013.

\bibitem{chong2018closing}
Frederic Chong.
\newblock Closing the gap between quantum algorithms and hardware through
  software-enabled vertical integration and co-design.
\newblock In {\em APS Meeting Abstracts}, 2018.

\bibitem{martonosi2019next}
Margaret Martonosi and Martin Roetteler.
\newblock Next steps in quantum computing: Computer science's role.
\newblock {\em arXiv preprint arXiv:1903.10541}, 2019.

\bibitem{huang2019alibaba}
Cupjin~Huang et~al.
\newblock Alibaba cloud quantum development platform: Applications to quantum
  algorithm design.
\newblock {\em arXiv preprint arXiv:1909.02559}, 2019.

\bibitem{fingerhuth2018open}
Mark Fingerhuth, Tom{\'a}{\v{s}} Babej, and Peter Wittek.
\newblock Open source software in quantum computing.
\newblock {\em PloS one}, 13(12):e0208561, 2018.

\bibitem{mckay2018qiskit}
David C~McKay et~al.
\newblock Qiskit backend specifications for openqasm and openpulse experiments.
\newblock {\em arXiv preprint arXiv:1809.03452}, 2018.

\bibitem{Broy.2006}
Manfred Broy.
\newblock Challenges in automotive software engineering.
\newblock In Leon~J. Osterweil, Dieter Rombach, and Mary~Lou Soffa, editors,
  {\em Proceeding of the 28th international conference on Software engineering
  - ICSE '06}, page~33, New York, New York, USA, 2006. {ACM Press}.

\bibitem{Bellman.1957}
Richard Bellman.
\newblock {\em Dynamic Programming}.
\newblock Dover Books on Computer Science. {Princeton University Press} and
  {Dover Publications}, Princeton, NJ, USA, 1 edition, 1957.

\bibitem{th_preskill_2019_nisq}
John Preskill.
\newblock Quantum computing in the nisq era and beyond.
\newblock {\em Quantum 2 (2018): 79.}, 2018.

\bibitem{Wright_2019}
K.~Wright et~al.
\newblock Benchmarking an 11-qubit quantum computer.
\newblock {\em Nature Communications}, 10(1), November 2019.

\bibitem{Pirandola_2018}
Stefano~Pirandola et~al.
\newblock Advances in photonic quantum sensing.
\newblock 2018.

\bibitem{Aramon_2019}
Maliheh~Aramon et~al.
\newblock Physics-inspired optimization for quadratic unconstrained problems
  using a digital annealer.
\newblock {\em Frontiers in Physics}, 7, April 2019.

\bibitem{Hitachi_2019}
Research and Ltd. Development~Group, Hitachi.
\newblock Research team led by the hitachi cambridge laboratory demonstrates an
  innovative hybdrid circuit for quantum computers, 2019.

\bibitem{Duda.2001}
Richard~O. Duda, Peter~E. Hart, and David~G. Stork.
\newblock {\em Pattern classification}.
\newblock Wiley-Interscience publication. Wiley, New York NY u.a., 2. ed.
  edition, 2001.

\bibitem{th_qiskit}
Gadi~Aleksandrowic et~al.
\newblock Qiskit: An open-source framework for quantum computing, 2019.

\bibitem{mcgeoch2019practical}
Catherine~C McGeoch, Richard Harris, Steven~P Reinhardt, and Paul~I Bunyk.
\newblock Practical annealing-based quantum computing.
\newblock {\em Computer}, 52(6):38--46, 2019.

\bibitem{th_pedregosa_2011_scikit-learn}
F.~Pedregosa et~al.
\newblock Scikit-learn: Machine learning in {P}ython.
\newblock {\em Journal of Machine Learning Research}, 12:2825--2830, 2011.

\bibitem{Bartkiewicz_2019}
Karol~Bartkiewicz et~al.
\newblock Experimental kernel-based quantum machine learning in finite feature
  space, 2019.

\bibitem{willsch2019support}
Dennis Willsch, Madita Willsch, Hans De~Raedt, and Kristel Michielsen.
\newblock Support vector machines on the d-wave quantum annealer.
\newblock {\em arXiv preprint arXiv:1906.06283}, 2019.

\bibitem{booth2016abstractions}
Michael Booth, Edward Dahl, Mark Furtney, and Steven~P Reinhardt.
\newblock Abstractions considered helpful: a tools architecture for quantum
  annealers.
\newblock In {\em 2016 IEEE High Performance Extreme Computing Conference
  (HPEC)}, pages 1--2. IEEE, 2016.

\bibitem{th_arute_2019_qs}
Frank~Arute et~al.
\newblock Quantum supremacy using a programmable superconducting processor.
\newblock {\em Nature 574, 505–510 (2019)}, 2019.

\end{thebibliography}

\end{document}